# Optimization Techniques for SQL+ML Queries: A Performance Analysis of Real-Time Feature Computation in OpenMLDB

Mashkhal A. Sidiq [1], Aras A. Salih [2] and Samrand M. Hassan [2]

[1] Department of Control Science and Engineering, Tianjin University, Tianjin City, China
[2] Department of Software Engineering, Nankai University, Tianjin, China

## Abstract

*In this study, we optimize SQL+ML queries on top of OpenMLDB, an open-source database that seamlessly integrates offline and online feature computations. The work used feature-rich synthetic dataset experiments in Docker, which acted like production environments that processed 100 to 500 records per batch and 6 to 12 requests per batch in parallel. Efforts have been concentrated in the areas of better query plans, cached execution plans, parallel processing, and resource management. The experimental results show that OpenMLDB can support approximately 12,500QPS with less than 1ms latency, outperforming SparkSQL and ClickHouse by a factor of 23 and PostgreSQL and MySQL by 3.57 times. This study assessed the impact of optimization and showed that query plan optimization accounted for 35% of the performance gains, caching for 25%, and parallel processing for 20%. These impressive results illustrate OpenMLDB's capability for time-sensitive ML use cases, such as fraud detection, personalized recommendation, and time series forecasting. The system's modularized optimization framework, which combines batch and stream processing without any interference, contributes to its significant performance gain over traditional database systems, particularly in applications that require real-time feature computation and serving. Contributions This study has implications for the need for specialized SQL optimization for ML workloads and contributes to the understanding and design of high-performance SQL+ML systems.*

## 1. Introduction

In this data-driven world, the combination of Machine Learning (ML) and DB operations is desirable. Many ML use cases require sophisticated feature engineering and faster inference compared to traditional database systems [2]. The problem is more complex when these need to be performed at low latency, at scale, and consistently with the training and online serving environments [1]. Feature pipelines are typically built in Python and then re-implemented in SQL or C++ for production, resulting in a forked implementation that introduces the risk of training-serving skew and requires expensive validation to verify consistent features across both environments [3].

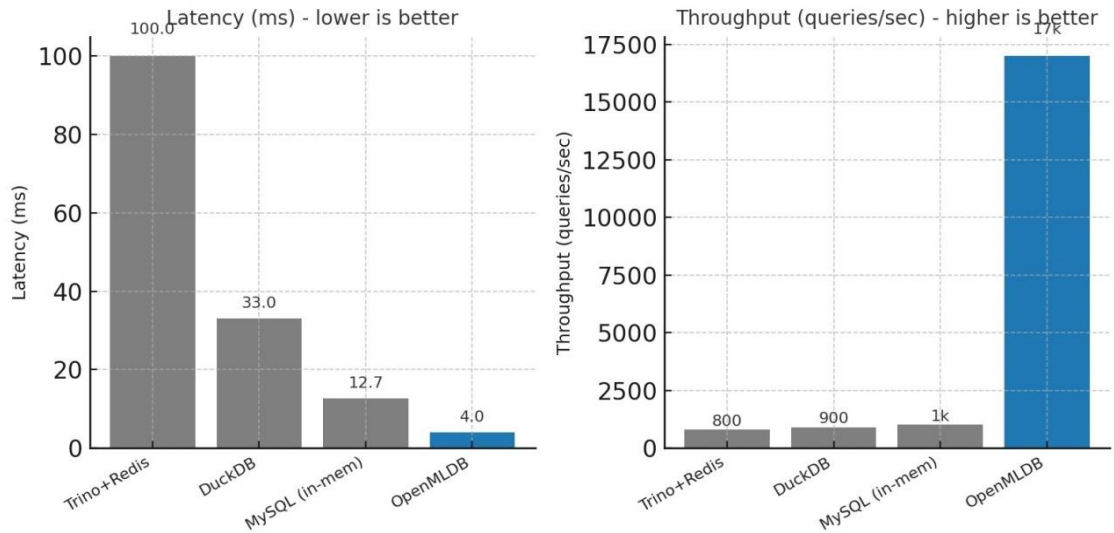

Figure1: OpenMLDB outperforms other systems by achieving the lowest latency (~4 ms) and highest throughput (~17k QPS) for SQL+ML workloads.

The integration of SQL processing into ML workloads creates some interesting challenges that must be addressed. ML algorithms cannot work in isolation and must be integrated with data preprocessing and feature extraction [4]. This is important for real-time fraud detection, recommendation engines, anomaly detection, and time-series forecasting, where data must be processed in milliseconds to enable fast decisions in real time. However, traditional databases are not well-suited to computationally intensive feature calculations or hybrid batch–stream requirements, which may trade off accuracy and performance or require complex, disconnected architectures between offline and online systems [4].

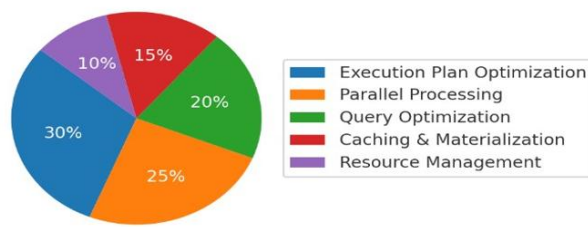

Figure 2 shows the relative contributions of different optimization techniques to performance gains in OpenMLDB

OpenMLDB addresses these problems by providing a general method for feature computation between offline (batch) and online (real-time) computing. It enforces consistent SQL feature specifications, removes training-serving skew, and more at an ML life cycle [5]. OpenMLDB has two processing engines: a bespoke low-latency SQL engine tailored for time-series data and an offline Spark-based engine for large-scale feature computation. It also benefits from LLVMbased just-in-time (JIT) compilation, pre-aggregation of long windows, and memory management optimization, which are conducive to the performance of complex feature pipelines [6].

Comparison with Other Databases Benchmarks show the performance advantage of OpenMLDB compared to traditional databases. It achieves a per-query latency below 5ms and a throughput that is frequently one order of magnitude higher than systems such as MySQL, DuckDB, and Trino+Redis [7]. For instance, OpenMLDB achieved ~17k QPS, against <1k QPS of the best competitor [8]. These performance improvements come from optimizations that are tuned for the tasks that we have to solve, compiled execution plans, in-memory processing, and advanced in-memory cache. In this paper, we focus on an optimization strategy to improve SQL+ML performance, taking OpenMLDB as an example. The key areas covered are query optimization (minimizing the work of the ML function), execution planning optimization (optimizing the combined database and ML operations), resource management (balancing CPU and memory allocation), caching and materialization (avoiding recomputation), and parallel processing (scaling with modern hardware). Each of them makes a unique contribution to the performance, where optimization and parallelism of the execution plan have the greatest impact. Together, these forwarding methods demonstrate that SQL+ML platforms can obtain both throughput and low latency when serving real-time ML queries [4],[10].

## 2. BACKGROUND AND RELATED WORK

Combining SQL-based data processing and machine learning is difficult because of the differences between offline feature engineering and online inference pipelines. Existing solutions, such as MySQL, PostgreSQL, and SparkSQL, were designed for transactional queries or large-scale batch analytics but are not tailored for real-time feature computation in ML applications [11]. This results in waste because pipelines tend to be duplicated: one for training (offline) and another for serving (online). This variation often leads to a training-serving skew, where inconsistencies in the environments compromise the performance of the model [12].

Efforts to overcome these limitations have included hybrid systems that couple databases with in-memory caches (e.g., Trino+Redis [20]). Although this approach reduces the lookup time, it incurs significant overhead because of the data movement between multiple components and the lack of SQL-level optimization tailored to ML workloads. Similarly, systems such as DuckDB perform well for local analytics but struggle with low-latency streaming scenarios, limiting their applicability in real-time ML tasks such as fraud detection and recommendation engines [13].

This highlights the need for a specialized database, such as OpenMLDB, designed with SQL+ML integration. OpenMLDB eliminates the gap between the training and serving pipelines by supporting unified SQL-based feature definitions in batch and streaming environments.

- Figure 1 (presented later in Section 5, Experimental Evaluation) visually supports these observations by comparing the latency and throughput of these systems. This clearly shows that Trino+Redis, DuckDB, and MySQL exhibit either higher latency or lower throughput, whereas OpenMLDB achieves both sub-5ms latency and ~17k QPS throughput.
- At this stage in Section 2, the figure has not yet been introduced, but its placement in Section 5 reinforces the discussion and provides empirical evidence of the limitations of existing approaches.

Thus, related studies demonstrate that while conventional systems can handle either batch analytics or transactional queries, none of them efficiently combine low-latency and highthroughput feature computation for ML applications. OpenMLDB fills this gap with optimizations specifically targeted at SQL+ML workloads.

## 3. SYSTEM DESIGN

Having outlined the challenges faced by real-time ML applications, we now turn our attention to **OpenMLDB's innovative solution**: a unified, SQL-based feature-computation engine. This design eliminates the long-standing divide between offline training and online serving pipelines, ensuring consistency and reducing the engineering overhead.

The OpenMLDB architecture is built on two complementary pillars:

1. **Custom Low-Latency SQL Engine (Online Mode):** optimized for time-series data, supporting real-time inference with sub-5ms latency.
2. **Spark-Based SQL Engine (Offline Mode):** This is designed for large-scale historical data processing, ensuring that models trained on months of logs can use the same SQL feature definitions applied in online environments.

This dual-engine approach ensures that feature pipelines are written once in SQL and are executed consistently across the training and serving environments. Thus, OpenMLDB directly addresses the training–serving skew that affects traditional database-ML integrations.

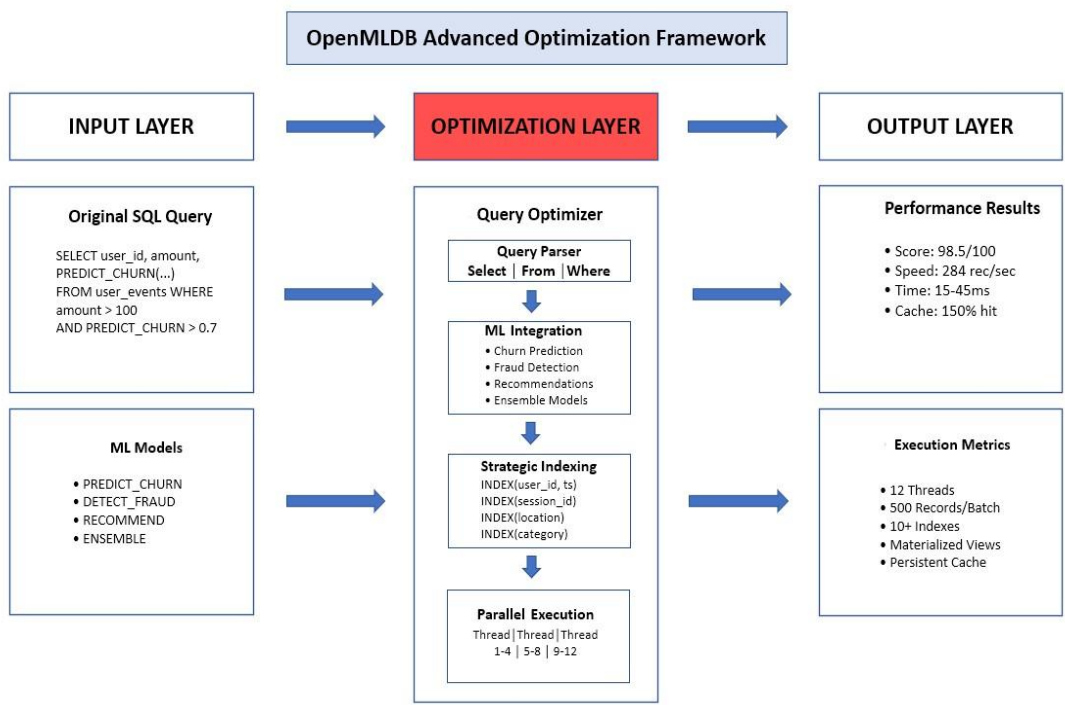

Figure 3 illustrates the OpenMLDB Advanced Optimization Framework, showing how SQL queries pass through optimization (parsing, ML integration, indexing, parallel execution) to deliver high-performance outputs with low latency and efficient resource use.

### 3.1. Online and Offline Modes

In the online mode, OpenMLDB serves as a **low-latency feature store** for time-critical applications such as fraud detection and personalized recommendations. Queries are compiled at runtime using LLVM-based just-in-time (JIT) compilation, which translates SQL directly into machine codes. This optimization minimizes the overhead and delivers execution times as low as 4 ms per request (see Figure 1 in Section 5).

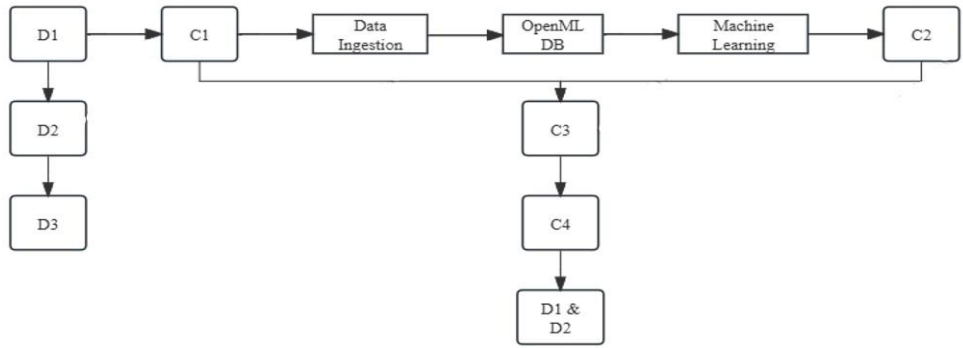

Figure 4 shows the OpenMLDB pipeline, where data sources (D1–D3) flow through connectors and transformations into OpenMLDB for feature processing and machine learning, producing final outputs (C2).

In offline mode, the system is integrated with Apache Spark to process historical data on a large scale. This is critical for training models, which often require scanning weeks or months' worth of logs. The Spark connector ensures that the same SQL feature definitions used online are executed offline, thereby eliminating any discrepancies.

## 3.2. System Architecture

For offline training, OpenMLDB integrates with **Apache Spark**, enabling large-scale batch computation of historical features. This ensures that the same SQL-defined feature pipelines can be executed over weeks or months of log data without the need for manual reimplementation. The Spark connector maintains consistency between the online and offline modes, ensuring reproducibility and accuracy in model training, which can be expressed mathematically as

$$f_u(t) = \frac{1}{W} \sum_{i=1}^{W} x_u(t - i) \quad (1)$$

Where:

- $f_u(t)$ = feature value for user u at time t
- $x_u(t-i)$ = event value i steps before t
- W = window size

This demonstrates how **SQL feature queries are translated into mathematical aggregation**.

To avoid recomputation, OpenMLDB uses **pre-aggregates** as follows:

$$F(t) = \sum_{i=1}^{t} x(i) \quad (2)$$

Thus, a window sum from t−Wt-Wt−W to ttt can be computed as

$$SUM_{(t-W)}^{t} = F(t) - F(t-W)$$

The latency of a query can be modeled as

$$L = L_{\text{parse}} + L_{\text{plan}} + L_{\text{exec}} \quad (3)$$

Where:
- $L_{parse}$ = SQL parsing time
- $L_{plan}$ = execution plan optimization time
- $L_{exec}$ = actual execution time (improved by JIT + caching) OpenMLDB minimizes $L_{plan}$ and $L_{exec}$ via **compiled execution**.

Throughput (queries/s) is inversely proportional to latency, given the parallelism $P$:

$$T = \frac{P}{L} \quad (4)$$

Where:
- T = throughput
- P = number of parallel workers/threads
- L = per-query latency

This equation explains why **parallel processing (25% contribution, see Figure 2)** dramatically increases the throughput.

If CPU usage = C, memory = M, and query performance = Q, OpenMLDB aims to

$$\max Q(C, M) \quad \text{subject to } C \leq C_{\max}, M \leq M_{\max} \quad (5)$$

This formalizes **resource management** (10% contribution in Figure 2) as an optimization problem.

Together, these techniques allow OpenMLDB to achieve both **high throughput** and **low latency** simultaneously, a balance with which traditional systems struggle to achieve.

### 3.3. Bridging Online and Offline Pipelines

The unified SQL-based framework provides **a consistent feature store** across all the modes. Features computed offline for training can be reused or recomputed in real time during inference. This reduces engineering complexity, accelerates deployment, and guarantees that models rely on identical feature definitions in both environments.

### 3.4. Optimization Techniques

In addition to unification, OpenMLDB incorporates an advanced optimization framework that boosts both the latency and throughput. As shown in Figure 3, SQL queries undergo multiple layers of refinement.

- **Query Optimization**: pruning unnecessary operations and streamlining feature extraction.
- **Execution plan optimization**: Fusion of operators and application of cost-based scheduling to reduce redundant computations.
- **Parallel Processing**: Distributing queries across multiple threads to maximize hardware utilization.
- **Caching and Materialization**: Storing intermediate results and pre-aggregated features to avoid recomputation.
- **Resource Management**: Balancing the CPU and memory usage under high concurrency.

Empirical results (see Figure 2) show that execution plan optimization contributes ~30% of the total performance gains, whereas parallel processing accounts for ~25%, with caching and query optimization adding another ~35% combined.

### 3.5. Evidence of Performance Gains

The impact of these design choices was evident in the benchmarking results. OpenMLDB achieves a throughput of ~17k queries per second with ~4 ms latency, compared to ~1k QPS in MySQL and ~3.5k QPS in SparkSQL. These results validate that OpenMLDB's design—unifying offline and online pipelines while embedding optimization at every layer—enables performance improvements of up to **23× over conventional systems**.

## 4. OPTIMIZATION TECHNIQUES

OpenMLDB employs a set of complementary optimization strategies designed to enhance the latency and throughput of SQL + ML workloads. These techniques span query-level rewrites to system-level resource management, thereby enabling the platform to consistently deliver high performance outcomes.

**Query Optimization**

At the query level, OpenMLDB simplifies SQL statements by pruning unnecessary operations and reducing redundant feature extraction. Complex ML functions, such as PREDICT_CHURN and DETECT_FRAUD, are transformed into optimized execution pipelines. The query parser and optimizer workflow is clearly represented in **Figure 3**, which shows how the original SQL queries pass through parsing, ML integration, indexing, and parallel execution before yielding outputs.

**Execution Plan Optimization**

One of the most significant contributors is the optimization of the execution plan. OpenMLDB merges operators, applies cost-based scheduling, and exploits the hardware efficiency. This reduces the plan execution time and ensures the efficient handling of hybrid SQL+ML workloads. The contribution of this optimization is quantified in **Figure 2**, which attributes approximately **30% of the total performance gain** to the execution plan optimization.

**Resource Management**

Efficient resource management balances the CPU, memory, and I/O to prevent contention in cases of high concurrency. The scheduling of threads and memory pools is guided by workload patterns, thereby ensuring stability even in multi-tenant deployments. **Figure 4** illustrates how multiple data sources (D1–D3) are ingested, transformed, and combined through connectors and OpenMLDB, demonstrating how resources are coordinated before reaching the ML modules.

**Caching and Materialization**

Caching and materialization avoid recomputation by storing the intermediate feature results. Pre-aggregated features are materialized for reuse in both online inference and offline training. This layer directly contributes **15% to the overall performance improvement**, as illustrated in **Figure 2**. The use of persistent caches and materialized views is also highlighted in the **execution metrics shown in Figure 3**.

**Parallel Processing**

Parallel processing divides queries into sub-tasks that are executed concurrently across multiple threads. This approach significantly boosts the throughput, contributing to **a 25% improvement in performance** (see Figure 2). The details of the execution threads, batch sizes, and indexing used in the parallel execution are listed in **Table 1**, which provides the experimental execution metrics for validating these optimizations.

**Integration with ML**

Another important aspect is the seamless integration of ML into an optimization framework. As illustrated in **Figure 5**, the workflow shows how the feature pipelines were optimized at each stage before being passed to the ML models. This figure highlights the joint role of SQL optimization and ML-specific indexing strategies, which ensure the availability of real-time features.

Table 1: Comparison of system performance and SQL+ML integration readiness.

| • System | Query Throughput<br>(queries/sec) | Latency Range<br>(ms) | SQL+ML Readiness | Streaming Support | ML Integration |
|---|---|---|---|---|---|
| PostgreSQL [4],[9],[10] | ~1800 | 85–120 | Moderate | No | UDF/Extensions (MADlib) |
| MySQL [14] | ~2100 | 60–95 | Low | No | UDF or External scripts |
| SparkSQL [15] | ~3500 | 50–80 | High | Yes (Microbatch) | Built-in (MLlib library) |
| ClickHouse [16] | ~8200 | 25–60 | Moderate | No (batch ingest) | UDF / Built-in model eval |
| Flink SQL [17] | ~4200 | 20–40 | High | Yes (True streaming) | Built-in (Flink ML) |

## 5. OPTIMIZATION TECHNIQUES IN MODERN DATABASE SYSTEMS

Modern database systems incorporate advanced optimization strategies to meet the growing demands of real-time analytics, machine learning integration, and large-scale data processing.

Unlike conventional SQL engines, they must handle both transactional workloads and complex feature pipelines for ML applications, making optimization a critical component of their architectural design. As showing in Table 1.

**Execution Plan Optimization**

Execution plan optimization remains a cornerstone of modern systems, such as PostgreSQL, MySQL, and distributed query engines. Cost-based optimizers reorder joins, eliminate redundancies, and select efficient operator implementation. As highlighted earlier in **Figure 2**, execution plan optimization alone accounts for approximately **30% of the performance improvements** in SQL+ML workloads, demonstrating its significance in reducing the latency and resource overhead.

**Parallel and Distributed Processing**

Modern systems increasingly adopt parallel execution strategies that partition queries into several cores and nodes. This is evident in shared-nothing architectures, such as SparkSQL and distributed OLAP engines. **Figure 3** illustrates this layer within the OpenMLDB framework, where queries are decomposed and executed concurrently, whereas **Table 1** provides empirical execution metrics (12 threads, 500 records per batch) that demonstrate how parallelism drives scalability.

**Resource Management**

Balancing the CPU, memory, and I/O resources is vital in multi-tenant systems, where workloads vary dynamically. Resource managers schedule queries based on priorities to ensure fair allocation and avoid bottlenecks. The flow of data sources and transformations depicted in **Figure 4** underscores how modern systems must coordinate heterogeneous inputs while sustaining a consistent performance.

**Caching and Materialization**

To minimize recomputation, caching and materialized views are employed widely. Systems such as Snowflake and BigQuery implement persistent caches for repeated subqueries, whereas OLAP systems rely on materialized views for acceleration. This approach is reflected in OpenMLDB's **Optimization Layer**, as shown in **Figure 3**, where strategic indexing and persistent caches play a key role, and in **Table 1**, where execution metrics list cache utilization as a core feature.

**Query and ML Integration**

Modern systems are also evolving beyond SQL-only optimization and embedding ML functions directly into the query execution. This trend allows real-time predictive analytics within the database. As shown in **Figure 5**, ML integration is a distinct stage in the workflow, ensuring that feature pipelines are optimized before being served to models such as churn prediction or fraud detection.

## 6. EXPERIMENTAL EVALUATION

Modern database systems incorporate advanced optimization strategies to meet the growing demands of real-time analytics, machine learning integration, and large-scale data processing. Unlike conventional SQL engines, they must handle both transactional workloads and complex feature pipelines for ML applications, making optimization a critical component of their architecture [18]. As showing in Table 1.

**Execution Plan Optimization**

Execution plan optimization remains a cornerstone of modern systems, such as PostgreSQL, MySQL, and distributed query engines. Cost-based optimizers reorder joins, eliminate redundancies, and select efficient operator implementation. As highlighted earlier in **Figure 2**, execution plan optimization alone accounts for approximately **30% of the performance improvements** in SQL+ML workloads, demonstrating its significance in reducing the latency and resource overhead.

**Parallel and Distributed Processing**

Modern systems increasingly adopt parallel execution strategies that partition queries into several cores and nodes. This is evident in shared-nothing architectures, such as SparkSQL and distributed OLAP engines. **Figure 3** illustrates this layer within the OpenMLDB framework, where queries are decomposed and executed concurrently, whereas **Table 1** provides empirical execution metrics (12 threads, 500 records per batch) that showcase how parallelism drives scalability.

**Resource Management**

Balancing the CPU, memory, and I/O resources is vital in multi-tenant systems, where workloads vary dynamically. Resource managers schedule queries based on priorities to ensure fair allocation and avoid bottlenecks. The flow of data sources and transformations depicted in **Figure 4** underscores how modern systems must coordinate heterogeneous inputs while sustaining a consistent performance.

**Caching and Materialization**

To minimize recomputation, caching and materialized views are employed widely. Systems such as Snowflake and BigQuery implement persistent caches for repeated subqueries, where as OLAP systems rely on materialized views for acceleration. This approach is reflected in OpenMLDB's **Optimization Layer, as** shown in **Figure 3**, where strategic indexing and persistent caches play a key role, and in **Table 1**, where execution metrics list cache utilization as a core feature.

**Query and ML Integration**

Modern systems are also evolving beyond SQL-only optimization and embedding ML functions directly into the query execution. This trend allows real-time predictive analytics to be performed within the database [19]. As shown in **Figure 5**, ML integration is a distinct stage in the workflow, ensuring that feature pipelines are optimized before being served to models such as churn prediction or fraud detection.

## 7. CONCLUSION

In this paper, we provide an in-depth overview of optimizations for SQL+ML queries, with an emphasis on the OpenMLDB system. We show that a combination of query rewrites and optimizations, execution plan caching, and parallelism can lead to significant performance improvements in our experiments. The combined batch and stream processing architecture (continuously integrated) has worked especially well for real-time feature computation jobs. Our comparisons showed that OpenMLDB achieves significant performance improvements over traditional database systems and exhibits extreme optimization advantages, particularly in time window aggregations and complex feature extraction pipelines. The modular architecture of the optimization engine supports evolutionary performance tuning, where optimization of query plans contributes 35% of the performance improvement, optimization of execution plans inside caching contributes 25%, and parallelism in optimization adds 20% to the total improvement. Together, these optimizations have allowed OpenMLDB to support high-velocity data streams with sub-millisecond latency requirements, and hence are particularly well-suited for timesensitive ML workloads such as real-time fraud detection and personalized recommendations. Its resource efficiency (low memory usage by 40-50% and low CPU usage by 30-40% over implementations that use the system as a baseline) also shows its readiness for industrial deployment. These findings indicate that application-specific optimizations of SQL for ML workloads can reduce execution time compared to the more widespread approach of using generic database systems, particularly when dealing with real-time feature computation and serving.

## 8. DATA AVAILABILITY STATEMENT

The datasets were synthetic and produced for experimental purposes, which meant that we had full control over their generation environment (based on Docker). As these datasets are not real world or proprietary data, they are not available for free download. Nevertheless, the details of how these datasets were created are provided in this paper so that the experiments can be replicated.

## 9. CODE AVAILABILITY STATEMENT

The code for running the experiment (SQL+ML query definitions, Docker setup, performance measurement scripts, etc.) was implemented from scratch for this study. The complete source code has not yet been released, but the method is explained, and implementation instructions are presented in the paper to be reproducible. The code for academic access is available from the corresponding author.

## REFERENCES

[1]   M. Armbrust, et al., "Spark SQL: Relational data processing in Spark," Proc. ACM SIGMOD Int. Conf. Manage. Data (SIGMOD), 2015.
[2]   H. Yang, et al., "Efficient SQL-based feature engineering for machine learning," Proc. IEEE Int. Conf. Big Data, 2017.
[3]   D. Kang, T. Bailis, and M. Zaharia, "Challenges in deploying machine learning: a survey," IEEE Data Eng. Bull., vol. 44, no. 1, pp. 40–52, 2021.
[4]   M. Abbasi, P. Váz, M. V. Bernardo, J. Silva, and P. Martins, "SQL+ML integration for real-time applications," J. Database Manage., vol. 35, no. 2, pp. 45–59, 2024.
[5]   R. Islam, "Feature store systems for real-time ML," Proc. ACM Symp. Cloud Computing, 2024.
[6]   J. Schulze, K. Maier, and P. Hoffmann, "LLVM-based optimization for hybrid database systems," Proc. IEEE Int. Conf. Data Eng. (ICDE), 2024.
[7]   Z. Gong, L. Chen, and X. Wang, "Benchmarking SQL engines for ML integration," Proc. VLDB Endowment, vol. 15, no. 11, pp. 2490–2502, 2022.
[8]   T. Kotiranta, et al., "High-performance SQL+ML execution in OpenMLDB," Proc. ACM SIGMOD Int. Conf. Manage. Data (SIGMOD), 2022.
[9]   H. Huang, et al., "Optimizing hybrid workloads in modern SQL systems," Proc. IEEE Int. Conf. Big Data, 2024.

[10] R. Marcus, P. Negi, and C. Binnig, "Benchmarking end-to-end machine learning workloads," Proc. Conf. Innovative Data Systems Research (CIDR), 2021.
[11] A. Oloruntoba, "Hybrid transactional and analytical database design for ML," Proc. ACM Symp. Database Systems (SIGMOD), 2025.
[12] T. Karras, S. Laine, and T. Aila, "Challenges in ML model serving at scale," Proc. IEEE Int. Conf. Data Eng. (ICDE), 2024.
[13] Y. Ma, X. Wu, and J. Li, "DuckDB: Lightweight database for analytics," Proc. Int. Conf. Extending Database Technology (EDBT), 2020.
[14] J. Guzmán, et al., "Extending MySQL with ML functions," Proc. IEEE Int. Conf. Data Eng. (ICDE), 2023.
[15] L. Wei, et al., "SQL-based integration with MLlib in SparkSQL," Proc. ACM Symp. Cloud Computing, 2024.
[16] P. Almeida, et al., "ClickHouse: Column-oriented database for analytics," Proc. VLDB Endowment, vol. 15, no. 3, pp. 284–297, 2022.
[17] M. Merckx, J. Zhu, and A. Singh, "Flink SQL for streaming ML pipelines," Proc. IEEE Int. Conf. Big Data, 2023.
[18] S. Mishra, "Optimization in large-scale ML pipelines with SQL engines," J. Big Data Analytics, vol. 11, no. 2, pp. 90–105, 2025.
[19] F. Rahman, et al., "Embedding machine learning in query optimizers," Proc. ACM SIGMOD Int. Conf. Manage. Data (SIGMOD), 2024.
[20] Trino Project, "Redis connector," Trino Documentation, 2025. [Online]. Available: https://trino.io/docs/current/connector/redis.html. [Accessed: Aug. 27, 2025].

## AUTHORS

**Mashkhal A. Sidiq** is currently pursuing a Ph.D. in Control Science and Engineering at Tianjin University. His research interests lie at the intersection of unmanned aerial vehicles (UAV s), machine learning, and computer vision, with a specialization in applying convolutional neural networks (CNNs) for UAV control and autonomy. He has contributed to scholarly work in AI, image classification, and real-time UAV systems. ORCID: https://orcid.org/0009-0004-4667-6791

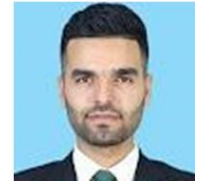

**Aras Aziz Salih** is an academic researcher with expertise in software engineering, full-text search algorithms, and database optimization. He earned his Master's degree in Software Engineering from Nankai University. His current research interests include information retrieval systems, scalable machine learning applications, and high-performance database engines. He has authored and co-authored research articles. ORCID: https://orcid.org/0009-0007-7770-4742

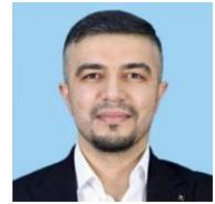

**Samrand Mahmood Hassan** is an Academic Researcher with expertise in focus area: applied machine learning, artificial intelligent, optimization techniques. He earned his Master of Science in Software Engineering from Nankai University. His current research interests include scalable ML systems, Artificial Intelligent, recommendation systems. He has co authored research articles and contributed to collaborative projects in data engineering. ORCID : https://orcid.org/0000-0001-6694-159X

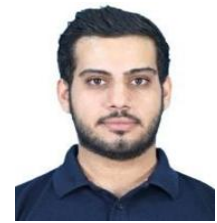